\documentstyle[aps,multicol,epsfig,prl]{revtex}

\input epsf

\def\prb{Phys.\ Rev.\ B}

\def\prl{Phys.\ Rev.\ Lett.\/}

\def\be{\begin{equation}}
\def\ee{\end{equation}}
\def\ba{\begin{eqnarray}}
\def\ea{\end{eqnarray}}

\def\LSCO{La$_{2-x}$Sr$_x$CuO$_4$}
\def\YBCO{YBa$_2$Cu$_3$O$_{7-\delta}$}

\def\BSCCO{Bi$_2$Sr$_2$CaCu$_2$O$_{8+\delta}$}
\def\C60{A$_x$C$_{60}$}
\def\LNSCO{La$_{1.25}$Nd$_{0.6}$Sr$_{0.15}$CuO$_{4}$}
\def\LNSCOx{La$_{1.4-x}$Nd$_{0.6}$Sr$_{x}$CuO$_{4}$}

\begin{document}

\title
{Evidence of Electron Fractionalization from Photoemission Spectra
in the High Temperature Superconductors}

\author{D.~Orgad$^1$, S.~A.~Kivelson$^1$, E.~W.~Carlson$^1$,
V.~J.~Emery$^2$, X.~J.~Zhou$^3$ and
Z.~X.~Shen$^3$}
\address
{1)  Dept. of Physics,
U.C.L.A.,
Los Angeles, CA  90095; 2)  Dept. of Physics, Brookhaven National Lab,
Upton, NY 11973;  3)  Dept. of Physics, Applied Physics and Stanford 
Synchrotron Radiation Laboratory, Stanford University, Stanford, CA 94305}

\date{\today}
\maketitle
\begin{abstract}


In the normal state 
of the high temperature superconductors \BSCCO\ and \LSCO, and in the related 
``stripe ordered'' material \LNSCO,
there is sharp structure in the measured single hole 
spectral function $A^<(\vec k,\omega)$ considered as a function of $\vec k$ 
at fixed small binding
energy $\omega$.
At the same time, as a 
function of $\omega$ at fixed $\vec k$ on much of the putative Fermi surface, 
any structure in $A^<(\vec k,\omega)$, other than the Fermi cutoff, is very
broad.  This is characteristic of the situation in which there 
are no stable excitations with the quantum numbers of the electron, 
as is the case in the one dimensional electron gas.
\end{abstract}

\begin{multicols}{2}
\narrowtext

In a Fermi liquid, the elementary excitations have the 
quantum numbers of an electron, so the one particle spectral function, 
$A(\vec k,\omega)$, is peaked at 
$\omega=\epsilon(\vec k)=\vec v_F(\vec k_F)\cdot (\vec k-\vec k_F)$,
where $\epsilon(\vec k)$ is the quasiparticle dispersion relation.
The single hole piece, $A^<(\vec k,\omega)$, 
can be measured 
using angle resolved photoemission spectroscopy (ARPES). 
The lifetime of
the quasiparticle, $\tau(\vec k)$, can be determined from the width of 
the peak in the ``energy distribution curve'' (EDC) defined by
considering
$A^<(\vec k,\omega)$ at fixed $\vec k$ as a function of $\omega$:
\be
1/\tau =\Delta\omega \; .
\ee

A check on the consistency of this picture can be obtained by studying 
the ``momentum distribution curve'' (MDC), {\it i.e.} by studying the 
width $\Delta k$ of the peak in $A^<(\vec k,\omega)$ at fixed binding 
energy, $\omega$.
So long as the quasiparticle excitation is well-defined, 
({\it i.e.} the decay rate is small compared to the binding energy), 
these two widths
are related by 
\be
\Delta\omega=v_F\Delta k \; ,
\label{widths}
\ee
where $v_F$ is the renormalized Fermi velocity which is directly measured. 
This well established Fermi-liquid
theoretic picture as applied to normal metals has
recently been observed in ARPES measurements of surface
states on Molybdenum by Valla {\it et al}.\cite{Mo}

The ways in which strong correlation effects can lead to the breakdown 
of Fermi liquid theory in more than one dimension are not well understood.
However, non-Fermi liquid behavior is generic in the theory of the one 
dimensional electron gas (1DEG), where  there are no elementary 
excitations with the quantum numbers of the hole \cite{review}.
Because of the celebrated separation of charge and spin, 
a hole (or an electron) is always 
unstable to decay into two or more elementary excitations, 
of which one or more carries its spin and 
one or more carries its charge. Consequently, $A^<(k,\omega)$ 
does not have a pole contribution, but rather consists of 
a multi-particle continuum. If both the spin 
and charge excitations are gapless, elementary kinematics implies 
that at $T=0$, $A^<(k,\omega)$ is nonzero only for negative frequencies
such that
\be
\label{kinematics}
|\omega| \ge{\rm min}(v_c,v_s)|k| \; ,
\ee
as shown in Fig. 1. (We define $A^<(k,\omega)$ as the Fourier 
transform of the hole piece of the single particle Green function with 
respect to $kx-\omega t$ and measure the wave vector and frequency 
relative to $k_F$ and $E_F$ respectively, so $-\omega$ is the electron 
binding energy.)

\begin{figure}[bht]
\narrowtext
\begin{center}
\leavevmode
\noindent
\hspace{0.3 in}
\centerline{\epsfxsize=2.6in \epsffile{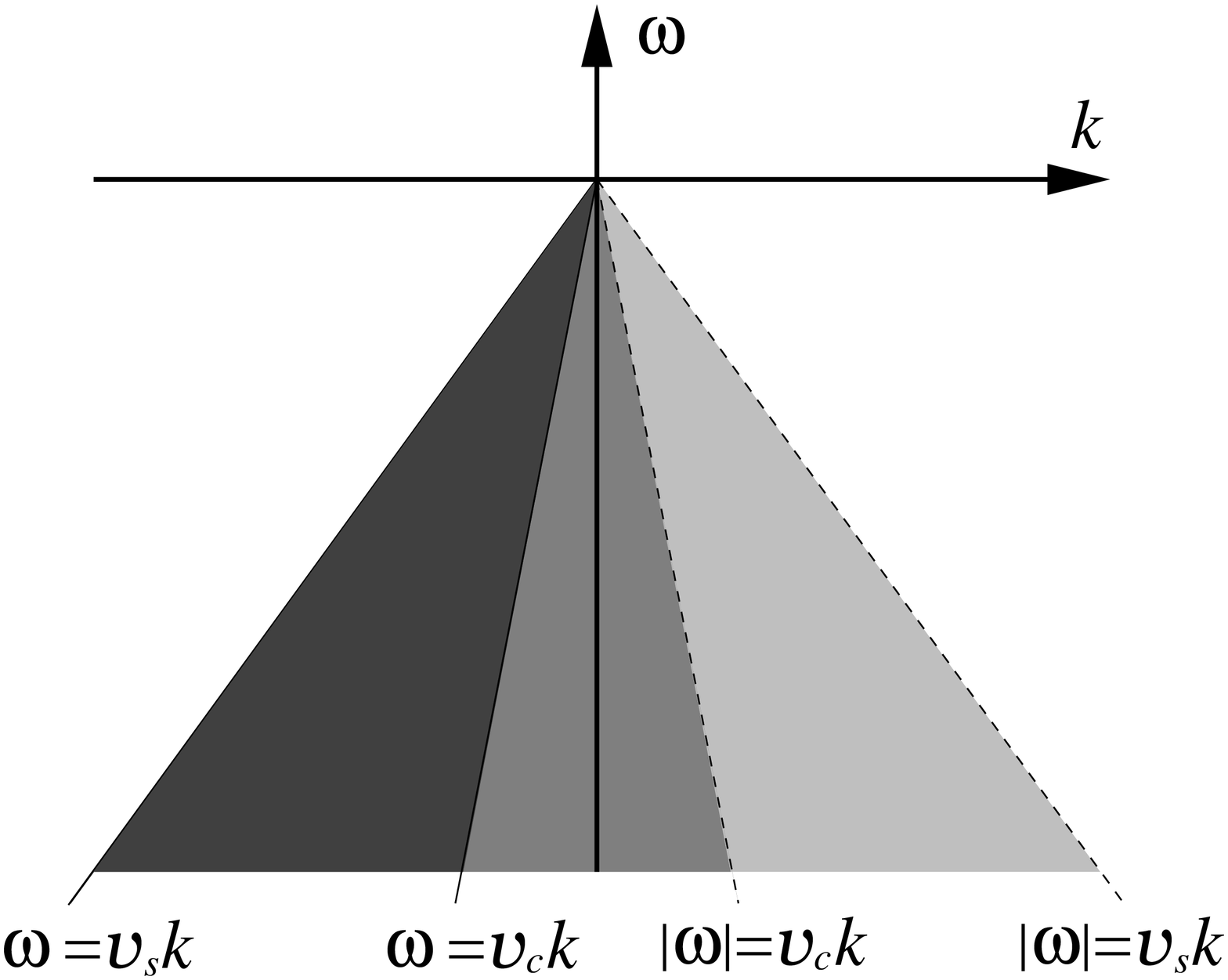}}
\end{center}
\caption
{Kinematic constraints:  $A^<(k,\omega)$ for the 1DEG is nonzero at $T=0$ only 
in
the shaded region of the $(k,\omega)$ plane. In the spin-rotationally invariant
case, $K_s=1$, $A^<(k,\omega)=0$ in the lightly shaded region, as well. If in
addition, $K_c=1$, $A^<(k,\omega)=0$ outside of the darkest region. Here 
$v_c>v_s$.}
\label{fig:1}
\end{figure}

Clearly, at $T=0$ and constant energy $\omega$ there will be nonzero spectral
weight in a region of $k$ of width $\Delta k = 2|\omega|/
{\rm min}(v_c,v_s)$, and a peak in the MDC with a full-width at half maximum
equal to some fraction of this.  
At finite temperature,
one effectively averages  over $\omega$ in a range $T$, giving rise to a
$\Delta k$ proportional  to the greater of $T$ and $|\omega|$.  By contrast, at
$k=0$, the shape of  the EDC 
is not given by the
kinematics at all, but is rather determined by the details of the matrix
elements linking the  one hole state to the various multi particle-hole states
which form the continuum.  In this case, the spectrum has a
non-universal power-law behavior with exponents determined by
the interactions in the 1DEG.


Figures 2 and 3 show finite temperature spectral functions of 
the 1DEG for various interaction strengths ({\it i.e.} for various
values of the charge Luttinger exponent $K_c$).
{\it It is a direct,  general, and dramatic
consequence of electron fractionalization that the MDC is much more highly
constrained  by kinematics than the EDC, which can often be very broad
compared to the MDC.} When such a dichotomy can be established experimentally,
we believe it represents strong evidence of electron fractionalization.
This dichotomy has been observed in
the measured spectral functions of
\LNSCO\ (LNSCO)\cite{shen1,shen2}  shown in Fig. 3, \LSCO\ 
(LSCO)\cite{shen2,ino1,ino2},
and slightly underdoped and even
optimally doped
\BSCCO\ (BSCCO)\cite{valla2,shen3}   
in the normal  state (at $T > T_c$). (For instance, see Fig. 2 of Ref.
\cite{valla2}.)

We would like to emphasize that spin-charge separation is sufficient 
but not necessary for fractionalization. The spinless 1DEG 
is an example where the electron decays into a multitude of left and 
right moving density waves. Here too the EDCs lack quasiparticle 
peaks and exhibit power law tails. Similar signatures are found when 
quasiparticles are strongly scattered by 2+1 or 3+1 dimensional quantum 
critical fluctuations\cite{subir} and in a ``marginal 
Fermi liquid''\cite{mfl} (MFL). However, in these last two cases the 
spectral weight is highly peaked at $\omega=v_Fk$ and 
Eq.~(\ref{widths}) holds to good approximation;  to logarithmic accuracy 
for the MFL.

Because the gapless Tomonaga-Luttinger liquid is a quantum critical system,
its response functions have a scaling form. Recently we have
obtained\cite{dror} explicit analytic expressions for these scaling 
functions under various conditions. In the spin rotationally invariant case,
the result for $A^<(\tilde k, \tilde\omega)$  at finite temperature 
in terms of the scaled variables $\tilde\omega=\omega/2\pi T$ and 
$\tilde k=v_s k/2\pi T$ is

\ba
\label{Alut}
\nonumber
A^<(\tilde k,\tilde\omega)&\propto& \int_{-\infty}^{\infty} 
dq\,h_{\gamma_c+\frac{1}{2}}
[\tilde\omega-\tilde k+(1+r)q] \\
&\times&h_{\gamma_c}[\tilde\omega-\tilde k-(1-r)q]
h_{\frac{1}{2}}(2\tilde k-2rq) \; ,
\ea
where $r=v_s/v_c$ is the ratio of the spin and charge velocities and 
$h$ is related to the Beta function, $B(x,y)$,
\be
\label{hk}
h_\gamma(k)={\rm Re}\left[(2i)^{\gamma}B\left(\frac{\gamma-i k}{2},1-\gamma
\right)\right] \; .
\ee
Here, we have introduced the critical exponents
\be
\label{gdef}
\gamma_\alpha=\frac{1}{8}(K_\alpha+K_\alpha^{-1}-2) \; ,
\ee
which are expressed in terms of the Luttinger parameters $K_{\alpha}$ with
$\alpha = c,s$ for charge and spin respectively. For a 
spin-rotationally invariant system $K_s$ approaches 1 at the fixed point. 
Therefore we have set $K_s = 1$.  

The kinematics discussed following Eq. (\ref{kinematics}) becomes evident 
once the $T \rightarrow 0$ limit of Eq. (\ref{Alut}) is considered, by using 
the asymptotic behavior $h_\gamma(|k|\rightarrow\infty)\propto (-k)^{\gamma-1}
\Theta(-k)$. 
An interesting subtlety occurs in the spin-rotationally invariant case, which
results in a more stringent constraint on the extent of the multi-particle
continuum than is implied by pure kinematics.  In this case, at the fixed 
point, the spin
correlators do not mix left and right moving pieces. As a consequence,
$A^<(T=0)$ vanishes if $v_s< v_c$ and $k>0$ when $\omega$ is in the 
range $v_s k\le |\omega|\le v_c k$, even if the kinematic constraints
in Eq. (\ref{kinematics}) are satisfied (see Fig. 1).

If $v_s<v_c$ and both $K_s=1$ and $K_c=1$, so that the charge piece also does
not mix left and right movers, $A^<(T=0)$ vanishes unless $k <0$ and $v_s 
|k|\le|\omega|\le v_c |k|$, as shown in Fig. 1.  There is, of course, 
no special reason why $K_c$ should be precisely equal to 1, but if the 
interactions are only moderately strong ({\it i.e.} $\gamma_c\lesssim 0.2$), 
most of the spectral weight is still concentrated in this region. In such a 
circumstance, even though the electron fractionalizes, so long as $v_c/v_s$ is
not too large, $\Delta\omega \sim (v_c-v_s)|k|$, and $\Delta k\sim 
[(v_c-v_s)/v_cv_s] |\omega|$ at $T=0$, and similar expressions with $T$ 
substituting for $k/v_F$ and $\omega$ respectively at elevated temperatures. 
Thus, the spectral function resembles that of a marginal Fermi liquid.

\begin{figure}
\narrowtext
\begin{center}
\leavevmode
\noindent
\hspace{0.3 in}
\centerline{\epsfxsize=3.15in \epsffile{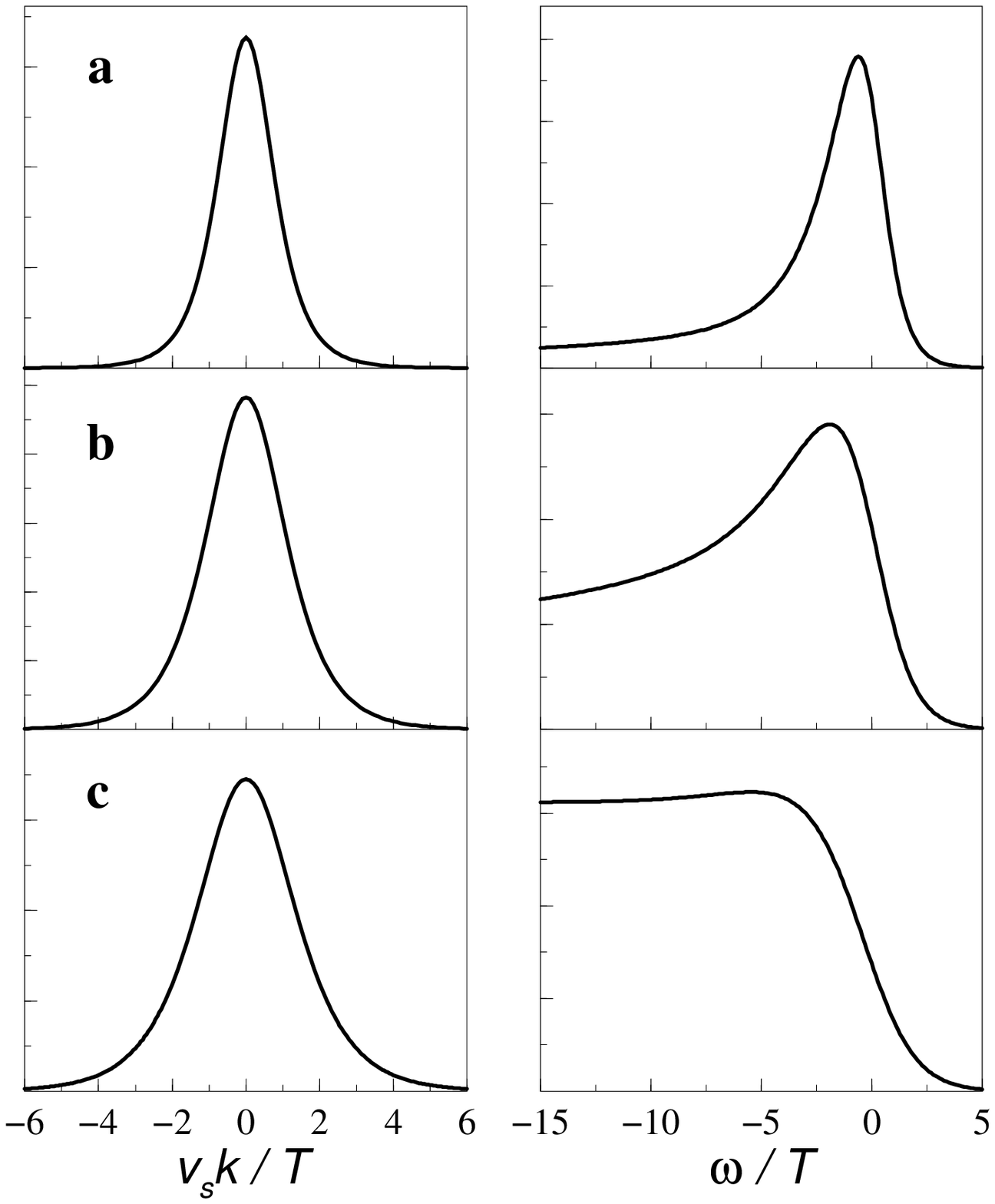}}
\end{center}
\caption
{MDCs for $\omega=0$ (left) and EDCs for $k=0$ (right) for the spin
rotationally invariant 1DEG   
with $v_c/v_s=3$ and a) $\gamma_c=0.1$, 
b) $\gamma_c=0.3$, and c) $\gamma_c=0.5$.}
\label{fig:2}
\end{figure}

In Fig.~2 we plot EDCs (for $k=0$) and MDCs (for $\omega=0$) generated 
using Eq. (\ref{Alut}) for $v_{s}/v_{c}\equiv r=1/3$ and various values 
of $\gamma_c$. 
The results depend only weakly on the choice of $r$.  However, 
as a function 
of $\gamma_c$, the EDCs change much more dramatically than do
the MDCs. In particular it is possible to eliminate any peak  
structure in the EDCs without broadening the MDCs substantially. 

It is also possible to obtain expressions for the spectral function of 
the spin gapped Luther-Emery liquid with $K_s=1/2$ \cite{dror}.  While
different in detail, the results in this case are grossly similar to those in
the gapless case aside from the fact that the Fermi edge is pushed back from 
the Fermi energy by the magnitude of the spin-gap.

The spectra in Fig. 2 are very reminiscent of the ARPES spectra seen
in the cuprates.  While the details vary from material to material, the
MDCs are always fairly sharp, while the EDCs broaden dramatically with
underdoping, especially in the $(\pi,0)$ region of the Brillouin zone (BZ).
{\it We take this as evidence of electron fractionalization in the normal 
state of these materials.}

There is experimental evidence that 
the physics of the 1DEG is also relevant to aspects of the electronic 
spectrum of the cuprates, especially at finite frequencies where local 
considerations are important.
\LNSCOx\ (x=0.1, 0.12, 0.15) is a ``stripe ordered'' 
non-superconducting relative of the high temperature superconductors 
\cite{tranquada}. The low energy ARPES spectral weight
is largely confined within patches in the ``anti-nodal'' region of the 
BZ with $k_x\approx\pm \pi/4$ and $k_y$ within 25\% of $\pi$, (and 
symmetry related regions of the BZ) \cite{shen1} consistent with the idea 
that the  spectrum is dominated by the 1DEG that lives along nearly 
quarter filled stripes. {\it Static} stripe order has been detected in 
\LSCO\ with $x<0.13$, but not for $x>0.13$,
or in BSCCO or \YBCO with $T_c\approx 90$K. However, in both
LSCO with $x>0.13$ and underdoped \YBCO, evidence of slowly fluctuating 
spin order has been detected with inelastic neutron
scattering\cite{fluctspin,fluctcharge}.
Evidence for fluctuating charge stripe order has also been 
reported\cite{fluctcharge}.  

The idea that the ARPES spectrum in the 
anti-nodal regions of BSCCO is dominated by quasi-one-dimensional physics 
has been discussed in a previous study \cite{carlson}.
Flat EDCs in the anti-nodal regions have been observed in stripe ordered
LNSCO\cite{shen1,shen2} and LSCO\cite{ino1,ino2}. Similarly in BSCCO, 
a dichotomy between a sharp peak in the MDC and the 
absence of a  peak in the corresponding EDC can be clearly seen in Fig. 3 
of Ref.\cite{valla2}. This dichotomy is most pronounced in the underdoped 
materials. The structures in the EDCs tend to sharpen with overdoping.    
Moreover, the spectral function shown in Fig. 1d of Ref.\cite{valla2} is 
consistent with our Fig. 1.

Electron fractionalization in the normal state of BSCCO near $(\pi,0)$ was 
inferred previously by us\cite{carlson} based on an independent, although 
somewhat less direct analysis. It has been observed that a sharp 
quasiparticle-like feature emerges in the superconducting state with 
$\Delta \omega \sim \Delta k/v_F$,  and a weight which is 
strongly temperature and doping dependent \cite{shen4,fedorov,shen5}. 
Empirically, it is observed that
the weight is roughly proportional to the superfluid density \cite{shen5}.
We have shown\cite{carlson} that this behavior can be understood as 
arising from a dimensional crossover from a one  dimensional 
(spin-charge separated) spectrum above $T_c$ to a two dimensional spectrum, 
consistent with the existence of an electron like quasiparticle, below $T_c$. 

\begin{figure}
\narrowtext
\setlength{\unitlength}{1in}
\begin{picture}(3.27,4.2)(0,-1.0)
\put(-0.05,-0.8){\epsfig{figure=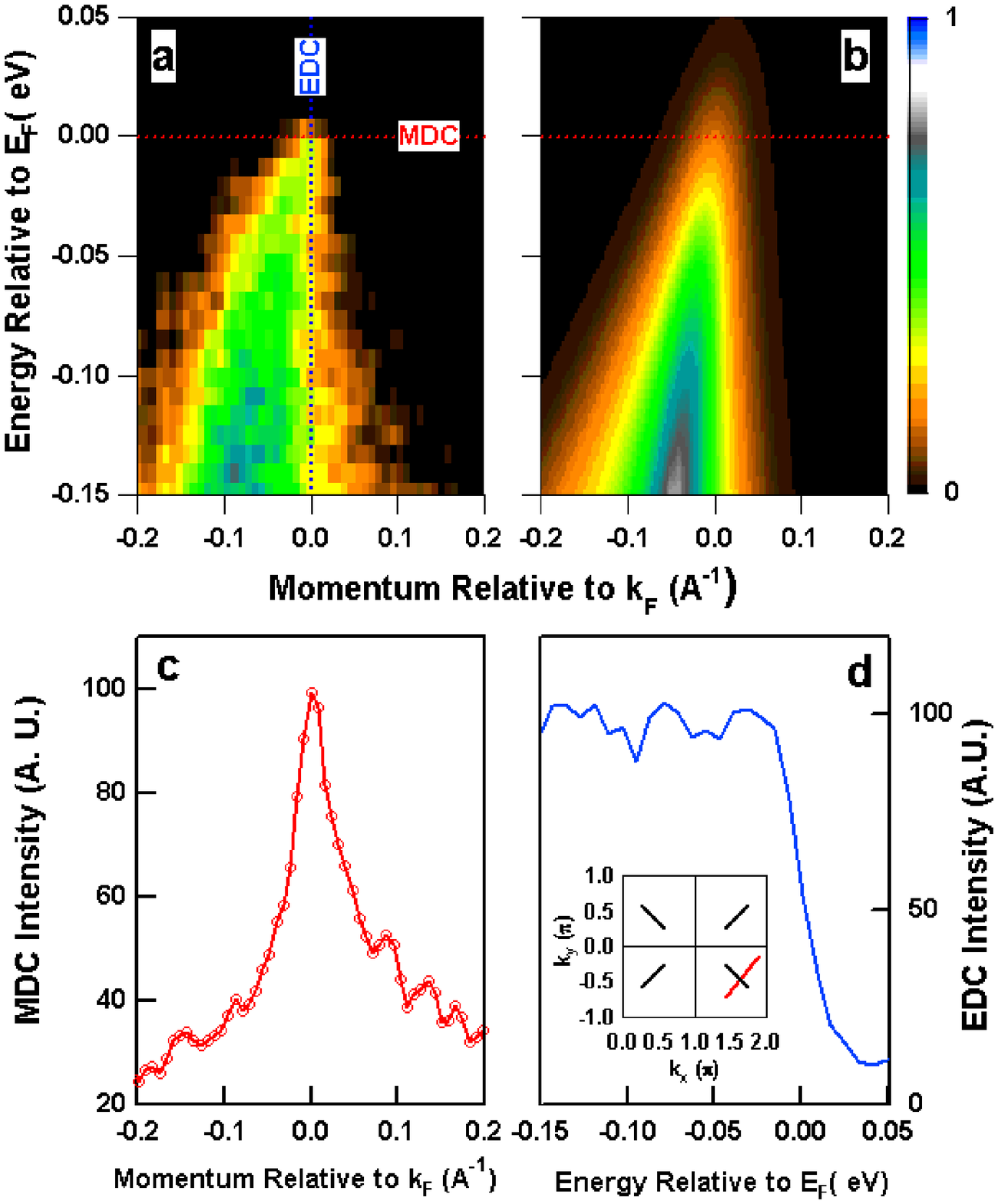,width=3.2in}}
\end{picture}
\caption
{Spectral functions of \LNSCO\  (experiment) and a spin-rotationally
invariant 1DEG (theory) with
$\gamma_c=0.5$, $v_s=0.7$eV-\AA, and $v_c=3.5$eV-\AA. 
In both cases, $T=$15K:  a) Experimental contour plot of $A^<(\vec k,\omega)$.
The data was collected along the line in the 
second BZ shown in red in the inset. The black diagonal lines 
indicate the position of the low energy Fermi segments; 
b) Contour plot for the 1DEG; 
c) Experimental MDC at $\omega=0$; 
d) Experimental EDC at $\vec k=(1.6\pi,-0.4\pi)$.}
\label{fig:3}
\end{figure}

Recently, it has been discovered\cite{shen2} that there is a second
component to the spectrum in LNSCO with small spectral weight 
in the ``nodal'' region, concentrated along straight Fermi segments
perpendicular to the $(0,0)$ to $(\pi,\pi)$ ray, as indicated in Fig.~3.  
Thus, the  distribution of low energy spectral weight in the BZ looks
qualitatively similar to that found in LSCO and BSCCO although in the latter
the nodal Fermi segments are considerably more curved. 
The origin of these nodal segments is not clear.
Possible sources for them may include a fluctuating 
stripe array\cite{salkola} and bond-centered stripes\cite{hanke}. 
It is also presently unknown whether such models lead to a 
fractionalized spectrum
in the nodal region, although for sufficiently flat Fermi segments 
and strong interactions this may be the case\cite{luther}. In the following 
we will take a heuristic point of view and also compare the nodal data 
with the 1DEG predictions.   

In Fig.~3 we show the one particle spectral function obtained in recent
experiments\cite{shen2} on LNSCO along a ray perpendicular to the Fermi
surface at the ``nodal point'' $\vec k_F=(1.6\pi,-0.4\pi)$ (a similar but 
weaker signal, due to matrix element effects, is observed along an 
equivalent cut in the first BZ). The measured spectra look similar to
those of the spin-rotationally invariant 1DEG with $\gamma_c=0.5$ 
(Figs. 2c and 3d)  
and spin and charge velocities $v_s\approx 0.7$eV-\AA and
$v_c\approx 3.5$eV-\AA. The value of $\gamma_c\sim 0.5$ corresponds to very
strong interactions.  
(In comparing theory with experiment, we confine ourselves to
the portion of the spectrum with binding energies less than  the
antiferromangetic exchange energy, $J \sim 0.1$eV.)


ARPES measurements along the nodal direction of  underdoped, optimally 
doped and overdoped LSCO \cite{shen2,ino1,ino2} reveal behavior of 
$A^<(\vec k,\omega)$  similar to that displayed in Fig. 3.
Normal state data from the nodal direction in
BSCCO is shown in Refs.\cite{valla2,nodal,valla1,argon}.
Here  there  is\cite{valla1} a well formed peak in the MDC, and a comparably
sharp peak in the EDC with $\Delta k \approx \Delta \omega/v_F$.  
If we compare the spectral
function with that of a 1DEG, we find that they look fairly 
similar provided we assume that $\gamma_c \approx 0.2$.
Of course, in this case it is also possible to imagine more
quasiparticle like interpretations of the data.

One such interpretation\cite{mfl} in terms of the marginal Fermi liquid
phenomenology is based on the recent observation\cite{valla2} that over 
most of the Fermi surface of BSCCO, the width of the MDC approximately 
satisfies the relation $\Delta k
\approx  (\Delta k)_o+(\Delta k)_1 T$ for a range of temperatures above $T_c$,
where $(\Delta k)_o$ depends on position along the Fermi surface but $(\Delta
k)_1$ does not.  However, in the same study it was found that, except for an
interval of Fermi surface near the nodal direction (comprising, perhaps, 30\%
of it), the EDC has little or no peak. Thus, taken at face value, the measured
widths of the EDCs outside this interval are {\it inconsistent} with a
marginal Fermi liquid form of the spectral function. We note that it
has been stressed for many years by Anderson\cite{anderson} and
Laughlin\cite{laughlin} that the breadth of the measured EDCs provides strong
evidence of electron fractionalization in the high temperature
superconductors;  the present analysis is similar in outline, although it
differs in many particulars.

While some aspects of the data admit to mundane explanations, such as
surface disorder, 
resolution effects where the dispersion is steep, and ambiguities due to any
background signal, the data set as a whole is more constraining.  
For example, while disorder could explain the breadth of the EDC near
$(\pi,0)$, this explanation is in apparent conflict with the sharpness of the
MDC, and, at least in BSCCO, with the emergence of a sharp peak in the EDC 
below $T_c$. Moreover, the structure in the EDC gets sharper with overdoping,
although the dispersion does not change substantially, and at least
in LSCO, the density of impurities increases.

We acknowledge stimulating discussions with
P.~Johnson, J.~Tranquada and T.~Valla.  
We thank S.~A.~Kellar, P.~V.~Bogdanov, A.~Lanzara, Z.~Hussain,
T.~Yoshida, A.~Fujimori, and S.~Uchida for collaboration.  SAK and EWC were
supported in part by NSF grant
No. DMR 98-14289.  VJE was supported in part by DOE grant No.
DE-AC02-98CH10886.  Data in Fig. 3 were collected at the Advanced Light
Source.  The SSRL work was supported by DOE grant No. DE-AC03-76SF00098.

\end{multicols}

\begin{references}

\vspace{-1 cm} 

\bibitem{Mo}  T.~Valla {\it et al.}, \prl\ {\bf 83}, 2085 (1999).  

\bibitem{review}  For a review, see V.~J.~Emery in 
{\it Highly Conducting One-Dimensional Solids}, eds. J.~T.~Devreese, 
R.~P.~Evrard, and V.~E.~van Doren,   (Plenum, New York, 1979).

\bibitem{shen1}  X.~J.~Zhou {\it et al.}, Science {\bf 286}, 268 (1999).

\bibitem{shen2}  X.~J.~Zhou {\it et al.}, cond-mat/0009002.
\label{ref:shen2}

\bibitem{ino1} A.~Ino {\it at al.}, Phys. Rev. B {\bf 62}, 4137 (2000).

\bibitem{ino2} A.~Ino {\it at al.}, cond-mat/0005370.

\bibitem{valla2}  T.~Valla {\it et al.}, Phys. Rev. Lett. {\bf 85}, 828 
(2000).
\label{ref:valla2}

\bibitem{shen3} A.~G.~Loeser {\it et al.}, Science {\bf 273}, 325 (1996);
H.~Ding {\it et al.}, Phys. Rev. Lett. {\bf 78}, 2628 (1997).

\bibitem{subir}  Explicit expressions for the spectral function in a quantum
critical system have been obtained by M.~Vojta, Y.~Zhang, and S.~Sachdev,
Phys. Rev. B {\bf 62}, 6721 (2000).  
See also S.~Caprara {\it et al.}, \prb\ {\bf 59}, 14980 (1999).

\bibitem{mfl} E.~Abrahams and C.~M.~Varma, cond-mat/0003135.

\bibitem{dror}  D.~Orgad, Phil. Mag. B {\bf 81}, 375 (2001).  This work
extends to finite temperature the earlier work of  
J.~Voit, Phys. Rev. B {\bf 47}, 6740 (1993) and V.~Meden and K.~Schonhammer,
Phys. Rev. B {\bf 46}, 15753 (1992).

\bibitem{tranquada}  J.~M.~Tranquada {\it et al.}, Nature (London) {\bf 375}, 
561 (1995).

\bibitem{fluctspin} J.~M.~Tranquada, Physica B {\bf 241}, 745 (1998);
H.~A.~Mook {\it et al.}, Nature (London) {\bf 395}, 580 (1998).

\bibitem{fluctcharge} H.~A.~Mook {\it et al.}, Nature (London) {\bf 404}, 
729 (2000).

\bibitem{carlson}  E.~W.~Carlson {\it et al.},
Phys. Rev. B {\bf 62}, 3422 (2000).



\bibitem{shen4}   A.~G.~Loeser {\it et al.}, \prb\ {\bf 56}, 14185 (1997).

\bibitem{fedorov}  A.~V.~Fedorov {\it et al.}, \prl\ {\bf 82}, 2179 (1999).

\bibitem{shen5}  D.~L.~Feng {\it et al.}, Science {\bf 289}, 277 
(2000); H.~ Ding {\it et al.} cond-mat/0006143.

\bibitem{salkola}  M.~Salkola {\it et al.}, \prl\ {\bf 77}, 155 (1996).

\bibitem{hanke} M.~G.~Zacher {\it et al.}, Phys. Rev. Lett. {\bf 85}, 
2585 (2000).

\bibitem{luther}  A.~Luther, \prb\ {\bf 50}, 11446 (1994).  See, also,
V.~J.~Emery {\it et al.}, \prl {\bf 85}, 2160 (2000).

\bibitem{nodal} C.~G.~Olson {\it et al.}, Science {\bf 245}, 731 (1989);
D.~S.~Dessau {\it et al.}, \prl\ {\bf 71}, 2781 (1993).
\label{ref:nodal}

\bibitem{valla1}  T.~Valla {\it et al.}, Science {\bf 285}, 2110 (1999).
\label{ref:valla1}

\bibitem{argon} A.~Kaminski {\it et al.}, Phys. Rev. Lett. {\bf 84}, 
1788 (2000); P.~V.~Bogdanov {\it et al.}, {\it ibid.} {\bf 85}, 2581 (2000).
\label{ref:argon}

\bibitem{anderson}  See, for example, P.~W.~Anderson, {\it The Theory of
Superconductivity in the Cuprates} (Princeton University Press, Princeton,
1997).

\bibitem{laughlin}  R.~B.~Laughlin, \prl\ {\bf 79}, 1726 (1997).


\end{references}
\end{document}